\documentstyle{article}
\begin{document}
\begin{titlepage}
\title{Constraints on Inflationary Solutions in the Presence of
Shear and Bulk Viscosity}
\author{
{\sc Henk van Elst $^{1}$, Peter Dunsby
$^{1,2}$ and Reza Tavakol $^{1}$}\\
\normalsize{1. {\it School of Mathematical Sciences, Queen Mary
and Westfield College,}}\\
\normalsize{{\it University of London, Mile End Road, E1 4NS, UK.}}\\
\normalsize{2. {\it Department of
Applied Mathematics, University of Cape Town,}}\\
\normalsize{{\it Rondebosch 7700, Cape Town, South Africa.}}}
\date{$\mbox{}$ \vspace*{0.3truecm} \\ \normalsize January 15 1994}
\maketitle
\thispagestyle{empty}
\begin{abstract}
Inflationary models and their claim to solve many of the
outstanding problems in cosmology have been the subject of a
great deal of
debate over the last few years. A major sticking point has been the
lack of both good observational and theoretical arguments to single
out one particular model out of the many that solve these problems.
Here we examine the degree of restrictiveness
on the dynamical relationship between the cosmological scale factor
and the inflation driving self-interaction potential of a minimally
coupled scalar field, imposed by the condition that the scalar
field is required to be real during a classical regime (the reality
condition). We systema\-tically look at the effects of this
constraint on many of the inflationary models found in the
literature within the FLRW framework, and also look at what happens
when physically motivated perturbations such as shear and bulk
viscosity are introduced. We find that in many cases, either the
models are totally excluded or the reality condition gives rise to
constraints on the scale factor and on the various parameters of
the model.
\end{abstract}
\vspace*{0.2truecm}
\begin{center}
{\it Subject headings:}\\ cosmology\,-\,galaxies: clustering,
formation\,-\,hydrodynamics\,-\,relativity
\end{center}
\end{titlepage}
\section{Introduction} \label{sec:intro}
An appealing feature of the dynamical interaction of a minimally
coupled homogeneous scalar field $\phi(t)$ with a self-interaction
potential $V(\phi)$ and the
Friedmann--Lema\^{\i}tre--Robertson--Walker (FLRW) models is that
the combination of the coupled Einstein--Scalar--Field
equations can be written in the form ($c=1$) \cite{ellismad}:
\begin{equation}
\label{V}
V\left[\,\phi(t)\,\right] = \frac{1}{\kappa_{0}}\,\left[ \,\dot{H}
+ 3\,H^{2} + 2\frac{k}{S^{2}}\, \right]\;,
\end{equation}
\begin{equation}
\label{phidot}
\dot{\phi}^{2}(t) = - \frac{2}{\kappa_{0}}\,\left[ \,\dot{H}
- \frac{k}{S^{2}}\, \right]\;,
\end{equation}
where $H:=\frac{\dot{S}}{S}$, $S$, $k$ and $\kappa_{0}$ denote
the Hubble parameter,
the cosmological scale factor, the curvature parameter and the
gravitational coupling constant respectively. Note that a solution
to these two equations will satisfy the remaining Klein--Gordon
equation automatically due to the vanishing of the covariant
divergence of the energy-momentum tensor. Equations (1) and (2)
can be treated as
a dynamical correspondence between the spaces of the functional
form of the scale factor $S(t)$, the corresponding scalar field
potential $V(\phi)$ and the space of relevant parameters such as
the curvature parameter and for example a shear coefficient. We
denote these spaces by $\cal{S}$, $\cal{V}$ and $\cal{P}$
respectively \cite{est}.
Consequently, given a functional form for the scale factor
together with a specification of the system parameters, one can
{\em in principle} find the corresponding potential and vice versa
\footnote{The inverse, however, is not as straightforward
since it involves explicitly solving the Klein--Gordon equation.
In practice, the problem is made manageable by employing
simplifying assumptions, such as the ``slow-rolling''
approximation or an exponential form for the potential.}.
The existence of
this correspondence might suggest that one is free to choose
any scale factor $S(t)$ together with a set of parameters and find
the corresponding $V(\phi)$. Now apart from the fact that in
practice this conversion may not be possible analytically, the
specific form of these equations set important constraints on the
concrete nature of this correspondence; namely: (i) the dynamical
relationship between
$S(t)$ and $V(\phi)$ may turn out to be fragile, in the sense that
small changes in $S(t)$ may result in qualitatively important
changes in the form of the corresponding $V(\phi)$ \cite{est} and
(ii)  since we confine ourselves to the classical regime where the
scalar field $\phi$ is supposed to be real, not all functional
forms of $S(t)$ may be permissible for
all values of parameters (over all ranges). This is made precise
through the {\em reality condition} \cite{est,ellismad} which
amounts to the right hand side of Eq. (\ref{phidot}) being
non-negative, i.e.
\begin{equation}
\label{cons}
-\left[\,\dot{H} - \frac{k}{S^{2}}\,\right] \geq 0.
\end{equation}
A simple example of this arises in the familiar exponential case,
$S(t) = A\,\exp\omega t$, for which condition (\ref{cons}) becomes
$k/S^{2} \geq 0$, implying that $k$ can only take values $k=0,~
k=+1$. As a result, the exponential solution is excluded in the
$k=-1$ FLRW case in presence of a scalar field, indicating clearly
that in presence of scalar fields, {\em not} every point in
$\cal{S}$ has a physically meaningful image in $\cal{V}$ for all
$\cal{P}$. This could be viewed as a form of {\em fragility}, in
the sense that small
changes in the parameters of the system can result in drastically
different prohibitive effects through the reality condition.

Now given the multiplicity of the functional forms of $S(t)$ which
have been employed to produce inflationary effects and the fact
that the precise nature of the potential $V(\phi)$ is not usually
known, it is clear that any constraints would be of great
value as well as having a physical consequence.

Our aims in this paper are as follows: (i) to take a systematic
look at the effects of the constraint (\ref{cons}) on the various
functional forms of $S(t)$ found in the literature within the FLRW
framework; and (ii) to look at the effects of introducing
physically motivated perturbations such as shear and bulk viscosity
on the constraints for the FLRW results.

The organisation of the paper is as follows: in Section
\ref{sec:reality} we look at the constraints imposed by the reality
condition within the familiar class of FLRW models. In Section
\ref{sec:shear} we study the effects
of allowing non-zero shear (anisotropic) perturbations by
considering the Bianchi models of Type--I and Type--V. Section
\ref{sec:bulk} contains a study of the effects of dissipation
(bulk viscosity) and in Section \ref{sec:all} we consider what
happens when shear and bulk viscosity are both present. We find it
informative to summarise our results in tabular form.

\section{Reality Condition within the FLRW Setting}
\label{sec:reality}
A number of Ans\"{a}tze have been employed for the scale factor
$S(t)$ to describe a possible phase of inflationary expansion
in the early Universe. These are almost exclusively based on the
assumption of
the existence of a scalar field minimally coupled to a background
FLRW spacetime manifold. In this section we take a variety of all
such examples of $S(t)$ within the FLRW framework and study the
effect of the reality condition (\ref{cons}) in each case as a
function of the system's parameters. The important point is that
the reality condition can, for certain values of the parameters,
exclude the solution under consideration in the relevant
inflationary epoch. It turns out that the main reason
why the reality condition is not so restrictive in the case of
inflationary models commonly discussed in the literature is that
$k$ is
usually taken to be zero in such models (which as we shall see
below corresponds to the least restrictive reality conditions). In
the following we shall classify the models according to the sign of
$\dot{H}$ and for completeness we give the analytic expressions for
the related scalar field potentials $V(\phi)$ in those cases
where they are known.

\subsection{Models of Inflation with $\dot{H} < 0$}
\label{subsec:models}
\subsubsection{Power-law expansion}  \label{ssc:plaw}
In power-law inflation (see Lucchin, Matarrese, 1985
\cite{lucmat}),
the scale factor takes the form:
\begin{equation}
\label{plaw}
S(t) = A\,t^{n} \hspace{1cm} A, n\ \mbox{const} > 0, \hspace{0.5cm}
n > 1,
\end{equation}
with the corresponding expression for $V(\phi)$ in the case of
$k=0$ given by:
\begin{equation}
V(\phi) = \frac{n}{\kappa_{0}}\,(3n-1)\,\exp\left[\,\pm\sqrt{\,
\frac{2\,\kappa_{0}}{n}\,}\,(\phi-\phi_{c})\,\right]\;,
\end{equation}
where $\phi_{c}$ is an integration constant. Note that for
$k\neq 0$ Eq. (\ref{phidot}) cannot be inverted analytically,
which is the reason why no explicit potential $V(\phi)$ is known in
those cases. The inverse relation for the cosmological proper time
variable $t$ obtained from Eq. (\ref{plaw}) is:
\begin{equation}
t(S) = \left(\frac{S}{A}\right)^{\frac{1}{n}}\;,
\end{equation}
which allows the reality condition (\ref{cons}) to be written in
the form:
\begin{equation}
n\,A^{\frac{2}{n}}\,S^{\frac{2n-2}{n}} \geq - k\;.
\end{equation}
Clearly this can only be satisfied for all ranges of $S,~n$ and
$A$, in the cases $k=+1$ and $k=0$. For $k=-1$ there is a lower
bound on $S$ depending upon the values of $A$ and $n$ given by:
\begin{equation}
S\geq \frac{1}{n^{\frac{n}{2n-2}} A^{\frac{1}{n-1}}}\;.
\end{equation}
As a result, early inflationary epochs are excluded in the $k=-1$
case.

\subsubsection{de Sitter expansion from a singularity}
\label{ssc:sinh}
The scale factor in this case is of the form (see Ellis and Madsen,
1991 \cite{ellismad}):
\begin{equation}
\label{sinh}
S(t) = A\,\sinh \omega t \hspace{1cm} A, \omega\
\mbox{const} > 0\;,
\end{equation}
with the corresponding potential given by:
\begin{equation}
V(\phi) = \frac{3\,\omega^{2}}{\kappa_{0}} + \frac{2}{\kappa_{0}}
\left(\omega^{2} + \frac{k}{A^{2}}\right)\,
\sinh^{2}\left(\,2\frac{\omega}{\sqrt{\,\frac{2}{\kappa_{0}}
\left(\omega^{2} + \frac{k}{A^{2}}\right)\,}}\left(\phi-\phi_{c}
\right)\,\right)\ ,
\end{equation}
where $\phi_{c}$ is an integration constant. The inverse of
Eq. (\ref{sinh}) can be written as:
\begin{equation}
t(S) = \frac{1}{\omega}\,\mbox{arsinh} \frac{S}{A}\;,
\end{equation}
which allows the reality condition to be written in the form:
\begin{equation}
\omega^{2}\,A^{2} \geq - k\;.
\end{equation}
Again this is satisfied for all ranges of values of $S$, $\omega$
and
$A$ in the cases $k=+1$ and $k=0$. For the case $k=-1$, $A$ and
$\omega$ are required to be constrained by the condition $\omega
\ge
\frac{1}{A}$.

\subsubsection{Intermediate de Sitter expansion from a singularity}
\label{ssc:intdS}
As a slight modification of the $\sinh$--expansion (\ref{sinh})
given above, we consider a scale factor with
the functional form:
\begin{equation}
S(t) = A\,\sinh(\,\omega\,t^{\epsilon}\,) \hspace{1cm} A,\omega,
\epsilon \ \mbox{const} > 0\;.
\end{equation}
By inverting this relation we find the expression for the
cosmological proper time variable $t$:
\begin{equation}
t(S) = \left(\frac{1}{\omega}\,\mbox{arsinh}\frac{S}{A}\right)
^{\frac{1}{\epsilon}}\;.
\end{equation}
It follows that the reality condition can be written as:
\begin{equation}
\label{realsinhe}
-\epsilon\left(\epsilon-1\right)\omega\,\frac{S\,\sqrt{A^2+S^2}}
{\left[\,\frac{1}{\omega}\,
\mbox{arsinh}\frac{S}{A}\,\right]^{\frac{2-\epsilon}{\epsilon}}}
+ \epsilon^2\omega^2A^2
\left[\,\frac{1}{\omega}\,\mbox{arsinh}\frac{S}{A}\,\right]
^{\frac{2\epsilon-2}{\epsilon}}\geq -k\;.
\end{equation}
In general, $\epsilon$ is allowed to take any positive value
whatsoever,
but it is particularly instructive to take a look at the behaviour
of the reality condition when $\epsilon$ takes values in the
neighbourhood of $1$. Suppose we take $\epsilon$-values
given by $\epsilon=1+\alpha$ and
$\epsilon=1-\alpha$ respectively, where $0<\alpha\ll1$ is fixed.
Now as $S\rightarrow 0$, then we qualitatively recover from
(\ref{realsinhe}) the
reality condition of the ordinary $\sinh$--expansion (12)
due to the smallness of $\alpha$ and hence the discussion
in the above section would go through for the different values
of $k$. However, for large
$S$ the behaviour
depends on the case chosen: for $\epsilon = 1+\alpha$,
the condition (\ref{realsinhe}) would necessitate that an
{\em upper bound} be placed on $S$ for all values of the curvature
parameter $k$. On the other hand, when $\epsilon=1-\alpha$,
condition
(\ref{realsinhe}) would be satisfied for $k=0$ and $k=+1$,
while for fixed values of the positive parameters it imposes a
{\em lower bound} on $S$ when $k=-1$. This change in
behaviour arises from the
fact that the first term in (\ref{realsinhe}), which is the
dominant one for large $S$,  changes sign as
$\epsilon$ passes from $\epsilon < 1$ to $\epsilon > 1$.
Concluding, this example demonstrates
how a small change in the model (here in the functional form of
the scale factor) can
qualitatively change the consequences of the reality condition and
hence the physical status of the model.

\subsubsection{Intermediate exponential expansion}
\label{ssc:iee}
The scale factor in this case is given by (see Barrow, 1990
\cite{bar1}):
\begin{equation}
\label{iee}
S(t) = A\,\exp \left(\,\omega\,t^{\epsilon}\,\right) \hspace{1cm}
A, \omega, \epsilon\ \mbox{const} > 0,\hspace{0.5cm}
0 < \epsilon < 1\;,
\end{equation}
with the corresponding potential in the case of $k=0$ of the form:
\begin{eqnarray}
V(\phi) = -\,\frac{\epsilon\,(\epsilon-1)}{\kappa_{0}}\,\omega\,
\left[\,\frac{\kappa_{0}\,\epsilon}{8\,(1-\epsilon)\,\omega}\,
(\phi-\phi_{c})\,\right]^{-\frac{2-\epsilon}{\epsilon}}\\
+ \frac{3\,\epsilon^{2}\,\omega^{2}}{\kappa_{0}}
\,\left[\,\frac{\kappa_{0}\,\epsilon}{8\,(1-\epsilon)\,\omega}\,
(\phi-\phi_{c})\,\right]^{-\frac{4-2\epsilon}{\epsilon}}\;,
\nonumber
\end{eqnarray}
where $\phi_{c}$ is an integration constant. Again the equation
(\ref{phidot}) cannot be inverted analytically for $k\neq 0$ .
The inverse of Eq. (\ref{iee}) can
be written as:
\begin{equation}
t(S) = \left(\,\frac{1}{\omega}\,\mbox{ln}\,\frac{S}{A}\,
\right)^{\frac{1}{\epsilon}}\;,
\end{equation}
which allows the reality condition (\ref{cons}) to be written as:
\begin{equation}
\label{rciee}
\epsilon\,(1-\epsilon)\,\omega\,\frac{S^{2}}{\left[\,
\frac{1}{\omega}\,\mbox{ln}\,\frac{S}{A}\,
\right]^{\frac{2-\epsilon}{\epsilon}}} \geq - k\;.
\end{equation}
Now for the cases of $k=0$ and $k=+1$, condition (\ref{rciee}) is
clearly
satisfied for all $S\ge A,~A,~\omega$ and $0<\epsilon <1$. The case
$k=-1$
is, however, much more complex and depends on the values of the
parameters. For example, for $\epsilon \rightarrow 1$, $S$ could
take all values down to its (by definition) lowest bound $A$, if
$\omega$ is large
enough to ensure that (\ref{rciee}) is satisfied. For small values
of $\omega$ the situation becomes much more complex as is also the
case for values of $\epsilon$ close to zero.

\subsection{Classical de Sitter Exponential Expansion
($\dot{H} = 0$)}
Originally this model was considered by Guth (1981) \cite{guth81}
(see also Ellis, Madsen, 1991 \cite{ellismad}) in the form:
\begin{equation}
\label{dSexp}
S(t) = A\,\exp \omega t \hspace{1cm} A, \omega\ \mbox{const} > 0\;,
\end{equation}
with the corresponding potential given by:
\begin{equation}
V(\phi) = \frac{3\,\omega^{2}}{\kappa_{0}}
+ \omega^{2}\,\left(\phi-\phi_{c}\right)^{2}\;,
\end{equation}
where $\phi_{c}$ is an integration constant.
The inverse for the variable $t$ is readily given by:
\begin{equation}
t(S) = \frac{1}{\omega}\,\mbox{ln}\,\frac{S}{A}\;,
\end{equation}
which results in a reality condition of the simple form:
\begin{equation}
k \geq 0\;.
\end{equation}
This relation is obviously satisfied for all $S, ~\omega, ~ A$ in
the cases of $k=+1$ and $k=0$. However, the $k=-1$ case is excluded
as we already mentioned in the introduction.

\subsection{Models of Inflation with $\dot{H} > 0$}
In the cases where $\dot{H}$ is positive it is immediately clear
that the reality condition (\ref{cons}) can {\em only}
be satisfied for a curvature parameter of value $k=+1$.

\subsubsection{de Sitter expansion without a singularity}
This case was considered by Ellis and Madsen, 1991 \cite{ellismad}
in the form:
\begin{equation}
S(t) = A\,\cosh \omega t\hspace{1cm} A, \omega\ \mbox{const} > 0\;,
\end{equation}
with the corresponding potential given by:
\begin{equation}
V(\phi) = \frac{3\,\omega^{2}}{\kappa_{0}} + \frac{2}{\kappa_{0}}
\left(-\omega^{2}+\frac{k}{A^{2}}\right)\,
\sin^{2}\left(\,2\frac{\omega}{\sqrt{\,\frac{2}{\kappa_{0}}
\left(-\omega^{2}
+\frac{k}{A^{2}}\right)\,}}\left(\phi-\phi_{c}\right)
\,\right)\ ,
\end{equation}
where $\phi_{c}$ is an integration constant.
The inverse relation for $t$ in this case is given by:
\begin{equation}
t(S) = \frac{1}{\omega}\,\mbox{arcosh}\frac{S}{A}\;,
\end{equation}
which allows the reality condition to be written in the form:
\begin{equation}
\omega^{2}\,A^{2} \leq k.
\end{equation}
Clearly this can only be satisfied for $k=+1$, if $\omega \le
\frac{1}{A}$.

\subsubsection{Super-exponential expansion}
\label{ssc:see}
The form of the scale factor in this case has emerged out of
``super-inflation'' scenarios in higher-dimensional theories of
gravity (see Shafi, Wetterich 1985 \cite{shawet} and Lucchin,
Matarrese 1985 \cite{lucmat2}) and is of the form:
\begin{equation}
S(t) = A\,\exp \left(\,\omega\,t^{n}\,\right) \hspace{1cm} A,
\omega, n\ \mbox{const} > 0 \hspace{0.5cm} n > 1.
\end{equation}
Note that with $k=+1$ Eq. (\ref{phidot}) cannot be inverted
analytically in order to derive an explicit expression for
$V(\phi)$.
The inverse of $S(t)$ in this case can be written as:
\begin{equation}
t(S) = \left(\,\frac{1}{\omega}\,\mbox{ln}\,\frac{S}{A}\,\right)
^{\frac{1}{n}}\;,
\end{equation}
which yields a reality condition of the form:
\begin{equation}
\label{see}
n\,(n-1)\,\omega\,S^{2}\,\left[\,\frac{1}{\omega}\,\mbox{ln}\,
\frac{S}{A}\,\right]^{\frac{n-2}{n}} \leq k\;.
\end{equation}
For a discussion of the above relation we divide the possible
values of the parameter $n$
into the three ranges (i) $1<n<2$, (ii) $n=2$ and (iii) $n>2$. As
pointed out before, $k=+1$ is the only curvature parameter allowed.
In the range (i), the situation is complicated but for small enough
values of $\omega$ and $A$ we find that a {\em lower} bound
needs to be imposed
on $S$,
as the term involving the logarithm in the denominator on the LHS
blows up as
$S\rightarrow A$. Also for fixed values of the parameters there
exists an {\em upper} bound on $S$ because $S^{2}$ in the numerator
of the LHS
grows faster than the logarithm term. In (ii) and (iii) $S$ needs
to be bounded from above only, the bound in (ii) being
$S \le \sqrt{\frac{1}{2\omega}}$.

\subsubsection{Double exponential expansion}
\label{ssc:dee}
This so called ``hyper-inflationary'' case was introduced by Barrow
in 1988 \cite{barrow88}  within string-driven inflationary
cosmological models where the scale factor $S(t)$ can assume the
double exponential form:
\begin{equation}
S(t) = A\,\exp\left[\,B\,\exp\omega t\,\right] \hspace{1cm} A, B,
\omega\ \mbox{const} > 0\;.
\end{equation}
Again, note that with $k\neq0$, Eq. (\ref{phidot}) cannot be
inverted analytically and therefore no explicit potential $V(\phi)$
can be given. Solving for the variable $t$ the inverse in this case
is given by:
\begin{equation}
t(S) = \frac{1}{\omega}\,\mbox{ln}\,\left(\,\frac{1}{B}\,\mbox{
ln}\,\frac{S}{A}\,\right)\;,
\end{equation}
which results in a reality condition of the form:
\begin{equation}
\omega^{2}\,S^{2}\,\mbox{ln}\,\frac{S}{A} \leq k\;.
\end{equation}
Subject to the condition $\omega^{2}\,A^{2}\,e^{2B}\,B\le 1$
there exists an upper bound for $S$ for $k=+1$. Note that even
then, depending on the parameters $A,\,B$ and
$\omega$, the range of values of $S$ that will satisfy the
reality condition might be extremely small.

\section{Reality Condition in Presence of Anisotropic
Perturbations}  \label{sec:shear}
We start with the Einstein--Scalar--Field equations in presence of
shear for cosmological models of Bianchi Type--I and Type--V in the
form:
\begin{equation}
\label{raysig}
\dot{H} = - H^{2} - \frac{\kappa_{0}}{6}\,(\rho_{\phi}+3\,p_{\phi})
- \frac{2}{3}\,\sigma^{2}\;,
\end{equation}
\begin{equation}
\label{friedsig}
H^{2} = \frac{\kappa_{0}}{3}\,\rho_{\phi} - \frac{k}{S^{2}}
+ \frac{1}{3}\,\sigma^{2}\;,
\end{equation}
\begin{equation}
\label{scaleq}
\ddot{\phi} + 3H\,\dot{\phi} + \frac{\partial V}{\partial \phi}
= 0\;,
\end{equation}
and
\begin{equation}
\label{sigeq}
\dot \sigma = -3\,H\,\sigma   \Longrightarrow \sigma^{2}
= \frac{\Sigma^{2}}{S^{6}}, \hspace{0.5cm}
\Sigma=\mbox{const}.
\end{equation}
For a discussion of the kinematical quantities for a scalar field
(such as shear etc.) see Madsen, 1988 \cite{mad88}.
Now treating the scalar field as a perfect fluid, the total energy
density and thermodynamical pressure are given by:
\begin{equation}
\label{rho}
\rho_{\phi} = \frac{1}{2}\,\dot{\phi}^{2}+V(\phi)\;,
\end{equation}
and
\begin{equation}
\label{pressure}
p_{\phi} = \frac{1}{2}\,\dot{\phi}^{2}-V(\phi)\;.
\end{equation}
Substituting these into the field equations (\ref{raysig}) and
(\ref{friedsig}) and
solving for $\dot{\phi}^{2}$ and $V(\phi)$ we obtain \cite{est}:
\begin{equation}
\label{dotphi2}
\dot{\phi}^{2}(t) = - \frac{2}{\kappa_{0}}\,\left[ \,\dot{H}
- \frac{k}{S^{2}} + \sigma^{2}\, \right]\;,
\end{equation}
\begin{equation}
\label{phi2}
V\left[\,\phi(t)\,\right] = \frac{1}{\kappa_{0}}\,\left[ \,\dot{H}
+ 3\,H^{2} + 2\frac{k}{S^{2}}\, \right]\;.
\end{equation}
We should stress that in the cases of cosmological models of
Bianchi Type--I and
Type--V considered here, the curvature parameter $k$ can
only take values $k=0$ and
$k=-1$ \cite{ellmac}. Furthermore, note that the form of $V(\phi)$
is not
effected by the presence of the shear term in the field equations
(\ref{raysig}) and (\ref{friedsig}). This will, however, change
once we include dissipation in the next section.
 The analogue of the reality condition (\ref{cons}) in this case
becomes (see Ellis, Skea and Tavakol, 1991 \cite{est}):
\begin{equation}
-\left[\,\dot{H} - \frac{k}{S^{2}} + \sigma^{2}\,\right] \geq 0\;.
\label{cond2}
\end{equation}
In the following we examine what happens to the results of the
previous section in presence of shear. We note that all
the $\dot H >0$ models are excluded here, since they only hold for
the $k=+1$ case. Also a de Sitter exponential expansion for
$k=0$ is prohibited by (\ref{cond2}).

\subsection{Power-law expansion}  
The reality condition (\ref{cond2}) can in this case be written as:
\begin{equation}
n\,A^{\frac{2}{n}}\,S^{\frac{2n-2}{n}} - \frac{\Sigma^{2}}{S^{4}}
\geq - k\;,
\end{equation}
which implies that in both cases of $k=0$ and $k=-1$ there is a
lower bound for $S$.
Comparison with the result for $k=0$ of Section \ref{ssc:plaw}
shows that
small non-zero $\Sigma$ perturbations have important consequences
for the restrictiveness of the reality condition.

\subsection{$\sinh$--expansion}  
The reality condition (\ref{cond2}) becomes:
\begin{equation}
\omega^{2}\,A^{2} - \frac{\Sigma^{2}}{S^{4}} \geq - k\;.
\end{equation}
This implies that in the case of $k=0$ one requires a lower bound
on $S$ given by $S > \sqrt{\frac{\Sigma}{\omega A}}$, whereas for
$k=-1$ we have the following constraints:
$S> (\frac{\Sigma^2}{\omega^2A^2-1})
^{\frac{1}{4}}$ and $\omega^2 A^2 -1 > 0$.

\subsection{Intermediate $\sinh$--expansion}  
The reality condition (\ref{cond2}) becomes:
\begin{equation}
\label{realsinhes}
-\epsilon\left(\epsilon-1\right)\omega\,\frac{S\,\sqrt{A^2+S^2}}
{\left[\,\frac{1}{\omega}\,
\mbox{arsinh}\frac{S}{A}\,\right]^{\frac{2-\epsilon}{\epsilon}}}
+ \epsilon^2\omega^2A^2
\left[\,\frac{1}{\omega}\,\mbox{arsinh}\frac{S}{A}\,\right]
^{\frac{2\epsilon-2}{\epsilon}} - \frac{\Sigma^{2}}{S^{4}}
\geq -k\;.
\end{equation}
The shear term in (\ref{realsinhes}) only becomes important as
$S\rightarrow 0$. Of course, compared to the results for
(\ref{realsinhe}), this implies quite a substantial change, because
the reality condition would now require a lower bound on $S$ for
both $k=0$ and $k=-1$. On the other hand, as $S$ increases the
shear term vanishes and the results inferred from (\ref{realsinhe})
would hold equally well here, implying that the changing $\epsilon$
in the neighbourhood of $\epsilon=1$ would produce a significant
effect.

\subsection{Intermediate exponential expansion}  
The reality condition (\ref{cond2}) in this case becomes:
\begin{equation}
\epsilon\,(1-\epsilon)\,\omega\,\frac{S^{2}}{\left[\,
\frac{1}{\omega}\,
\mbox{ln}\,\frac{S}{A}\,\right]^{\frac{2-\epsilon}{\epsilon}}}
- \frac{\Sigma^{2}}{S^{4}} \geq - k\;.
\end{equation}
Again the overall situation is very complicated and depends on
values
of the free parameters $\epsilon,~\omega,~A$ and $\Sigma$.
However, for $\epsilon \rightarrow 1$ and for sufficiently large
$\omega$, similar results to those of Section \ref{ssc:iee} hold
and at least in such settings
those results are robust under non-zero perturbations of $\Sigma$.

\section{In Presence of Bulk Viscosity} \label{sec:bulk}
To study the effects of dissipation on the constraints imposed by
the reality condition in the FLRW cases, we look at the effects of
adding a non-zero bulk viscosity term. Now since using
the causal formalism of ``extended irreversible thermodynamics''
according to Israel \cite{isr} turns out to be intractable for our
purposes (as gradients of the potential $V(\phi)$ arise), we
restrict ourselves to a simplified model where we define an
effective
isotropic pressure $P_{eff}$ as a function of the total energy
density (\ref{rho}) for the scalar field, following Eckart's
first-order non-equilibrium thermodynamic theory
\cite{eck,ellis73}, such that:
\begin{equation}
\label{peff}
P_{eff}(\phi, \dot{\phi}) := \frac{1}{2}\,\dot{\phi}^{2}-V(\phi)
- 3\,\zeta\,H(\phi, \dot{\phi})\;,
\end{equation}
where we take the {\em coefficient of bulk viscosity} $\zeta > 0$
to be a constant (since otherwise, apart from the linear case,
taking $\zeta$ as a power-law function of the total energy density
(\ref{rho}) would result in cross terms between $\dot{\phi}^{2}$
and $V(\phi)$\,). Inserting from Eq. (\ref{peff}) into the
Einstein--Scalar--Field equations of the FLRW case,
we can again solve for
$\dot{\phi}^{2}$ and $V[\,\phi(t)\,]$ to obtain:
\begin{equation}
\label{dotphi3}
\dot{\phi}^{2}(t) = - \frac{2}{\kappa_{0}}\,\left[ \,\dot{H}
- \frac{k}{S^{2}} - \frac{3}{2}\,\kappa_{0}\,\zeta\,H\, \right]\;,
\end{equation}
and
\begin{equation}
\label{phi4}
V\left[\,\phi(t)\,\right] = \frac{1}{\kappa_{0}}\,\left[ \,\dot{H}
+ 3\,H^{2} + 2\frac{k}{S^{2}} - \frac{3}{2}\,\kappa_{0}\,\zeta\,H\,
\right]\;.
\end{equation}
The reality condition then becomes:
\begin{equation}
\label{cond3}
- \left[ \,\dot{H} - \frac{k}{S^{2}}
- \frac{3}{2}\,\kappa_{0}\,\zeta\,H\, \right] \geq 0\;.
\end{equation}

\subsection{Power-law expansion}  
In this case the reality condition (\ref{cond3}) becomes:
\begin{equation}
n\left[\,\left(\frac{A}{S}\right)^{\frac{1}{n}}
+ \frac{3}{2}\,\kappa_{0}\,\zeta\,\right]\left(\frac{A}{S}\right)
^{\frac{1}{n}}\,S^{2} \geq - k\;.
\end{equation}
This case is very similar to the one discussed in Section
\ref{ssc:plaw}. For $k=0,~k=+1$, the above condition is satisfied
for all values of the parameters, whereas for $k=-1$, there is a
lower bound for $S$.

\subsection{$\sinh$--expansion}  
The reality condition (\ref{cond3}) becomes:
\begin{equation}
\omega^{2}\,A^{2} + \frac{3}{2}\,\kappa_{0}\,\zeta\,
\omega\,S\,\sqrt{S^{2}+A^{2}} \geq - k\;.
\end{equation}
This case remains qualitatively the same as the related one
considered in Section \ref{ssc:sinh} so that the discussion given
there applies here equally well.

\subsection{Intermediate $\sinh$--expansion}  
The reality condition (\ref{cond3}) becomes:
\begin{eqnarray}
\label{realsinhebv}
& &-\epsilon\left(\epsilon-1\right)\omega\,\frac{S\,\sqrt{A^2+S^2}}
{\left[\,\frac{1}{\omega}\,
\mbox{arsinh}\frac{S}{A}\,\right]^{\frac{2-\epsilon}{\epsilon}}}
+ \epsilon^2\omega^2A^2
\left[\,\frac{1}{\omega}\,\mbox{arsinh}\frac{S}{A}\,\right]
^{\frac{2\epsilon-2}{\epsilon}}\\
& &+ \frac{3}{2}\,\kappa_{0}\,\zeta\,\epsilon\,\omega\,S\,
\sqrt{A^{2}+S^{2}}\,\left[\,\frac{1}{\omega}\,
\mbox{arsinh}\frac{S}{A}\,\right]^{\frac{\epsilon-1}{\epsilon}}
\geq -k\;.\nonumber
\end{eqnarray}
Pursuing the discussion analogously to the related case in Section
\ref{ssc:intdS}, we concentrate on a neighbourhood around
$\epsilon=1$. As $S\rightarrow 0$ the results of (\ref{realsinhe})
are also true here. However, due to the presence of the bulk
viscosity term, which turns out to be the dominant one for
large enough values of $S$, (\ref{realsinhebv}) is no longer
sensitive to the
change of sign of its first term as $\epsilon$ goes from values
$\epsilon<1$ to values $\epsilon>1$. Consequently, for $k=0$ and
$k=+1$ cases, (\ref{realsinhebv}) holds for large enough $S$
and one only has to put a lower
bound on $S$ when $k=-1$.

\subsection{Intermediate exponential expansion}  
The reality condition (\ref{cond3}) becomes:
\begin{equation}
\epsilon\,\omega\left[\,\frac{1-\epsilon}{\left[\,\frac{1}{\omega}
\,\mbox{ln}\,\frac{S}{A}\,\right]^{\frac{2-\epsilon}{\epsilon}}}
+ \frac{3}{2}\,\kappa_{0}\,\zeta\,\frac{1}{\left[\,\frac{1}{\omega}
\,\mbox{ln}\,\frac{S}{A}\,\right]^{\frac{1-\epsilon}{\epsilon}}}
\right]\,S^{2} \geq - k\;.
\end{equation}
For $k=0$ and $k=+1$ this condition is satisfied for all values
$S\ge A$. For $k=-1$
the overall situation is again very complicated and depends on
values
of the free parameters $\epsilon,~\omega,~A$ and $\zeta$.
However, for the case of $\epsilon \rightarrow 1$ and sufficiently
large
$\omega$ discussed in Section \ref{ssc:iee}, non-zero $\zeta$
perturbations do not produce any qualitative changes.

\subsection{de Sitter expansion}  
The reality condition (\ref{cond3}) becomes:
\begin{equation}
\label{bvdS}
\frac{3}{2}\,\kappa_{0}\,\zeta\,\omega\,S^{2} \geq - k\;.
\end{equation}
In the cases of $k=0$ and $k=+1$, this is satisfied for all values of the
parameters, but there are restrictions in the $k=-1$ case.
In particular, when $S\rightarrow A$ the condition (\ref{bvdS})
is satisfied for all values of $S$ defined by (\ref{dSexp}) only if
$A \geq \sqrt{\frac{2}{3\kappa_{0}\,\zeta\,\omega}}$,
otherwise there is a lower bound on $S$ with $S>A$. Note
that in contrast to the case with $\zeta=0$, where $k=-1$
is ruled out, any non-zero bulk viscosity perturbations would make
the $k=-1$ case (conditionally) permissible. This
will also be a feature of the next three examples and in this sense
all these examples are fragile with respect to such perturbations.

\subsection{$\cosh$--expansion}  
The reality condition (\ref{cond3}) becomes:
\begin{equation}
\omega^{2}\,A^{2} - \frac{3}{2}\,\kappa_{0}\,\zeta\,
\omega\,S\,\sqrt{S^{2}-A^{2}} \leq k\;.
\end{equation}
As $S\rightarrow A$ there is a lower bound for $S$ with $S>A$
when $k=0,~-1$. If $k=+1$, we
require that $\omega^2 A^2 \le 1$. Otherwise, there will be a lower
bound for $S$ with $S>A$ as well.

\subsection{Super-exponential expansion}  
The reality condition (\ref{cond3}) becomes:
\begin{equation}
\label{seecribv}
n\,\omega\left[\,(n-1)\,\left[\,\frac{1}{\omega}\,\mbox{ln}\,
\frac{S}{A}\,\right]^{\frac{n-2}{n}} - \frac{3}{2}\,\kappa_{0}\,
\zeta
\,\left[\,\frac{1}{\omega}\,\mbox{ln}\,\frac{S}{A}\,
\right]^{\frac{n-1}{n}}\,\right]\,S^{2} \leq k\;.
\end{equation}
Again, as in the discussion of the pure FLRW case in Section
\ref{ssc:see}, we divide the values of the parameter $n$
into the ranges (i) $1<n<2$, (ii) $n=2$ and (iii) $n>2$. For the
case
(i) the first term in (\ref{seecribv}) which is positive
 will dominate for small enough $\omega$ and $A$ and
therefore for the reality condition to hold for all $k$, a lower
bound ($S>A$) needs to be imposed on $S$. In (ii), a lower bound
on $S\,(S>A)$ is required for $k=0$ and $k=-1$, whereas for $k=+1$
we find the constraint $\omega \leq 1/2$. Finally, in (iii)
depending upon the values of $\kappa_{0}$ and $\zeta$, there
exists a lower
bound on $S\,(S>A)$ only for $k=-1$. For large values of $S$
the behaviour is complicated and will depend on the magnitude
of the product
$\kappa_{0}\,\zeta$. Note that the presence of the bulk
viscosity term has removed the necessity for an upper bound on $S$
for $k=+1$ case, as compared to the respective results in Section
\ref{ssc:see}.

\subsection{Double exponential expansion}  
The reality condition (\ref{cond3}) becomes:
\begin{equation}
\omega\left(\omega-\frac{3}{2}\,\kappa_{0}\,\zeta\right)\,S^{2}\,
\mbox{ln}\,\frac{S}{A} \leq k.
\end{equation}
For $k=0, ~+1$, the above relation is satisfied subject to $\omega
< 3/2\,\kappa_0\,\zeta$. With this restriction on $\omega$ one then
finds that there is a lower bound
on $S$ for $k=-1$ case. On the other hand for
$\omega > 3/2\,\kappa_{0}\,\zeta$,
$k=0$ and $k=-1$ are excluded and an upper bound needs to be put on
$S$ for $k=+1$, which depends on the values of the free parameters and
subject to the same conditions as given in Section \ref{ssc:dee}.

\section{Simultaneous Shear and Bulk Viscosity Perturbations}
\label{sec:all}
In this section we study the effects of ``switching on'' shear and
bulk viscosity perturbations simultaneously. The relevant equations
can be derived from the field equations (\ref{raysig}) and
(\ref{friedsig}) together
with Eq. (\ref{peff}) and can be solved for $\dot{\phi}^{2}$ and
$V[\,\phi(t)\,]$ to give:
\begin{equation}
\label{dotphi4}
\dot{\phi}^{2}(t) = - \frac{2}{\kappa_{0}}\,\left[ \,\dot{H}
- \frac{k}{S^{2}}
+ \sigma^{2} - \frac{3}{2}\,\kappa_{0}\,\zeta\,H\, \right]\;,
\end{equation}
and
\begin{equation}
\label{phi3}
V\left[\,\phi(t)\,\right] = \frac{1}{\kappa_{0}}\,\left[ \,\dot{H}
+ 3\,H^{2} + 2\frac{k}{S^{2}}
- \frac{3}{2}\,\kappa_{0}\,\zeta\,H\, \right]\;.
\end{equation}
The reality condition then becomes:
\begin{equation}
\label{cond4}
- \left[ \,\dot{H} - \frac{k}{S^{2}} + \sigma^{2}
- \frac{3}{2}\,\kappa_{0}\,\zeta\,H\, \right] \geq 0\;.
\end{equation}
Again we recall
that, due to the presence of the shear term, the curvature
parameter $k$ can only take the values $k=0$ and $k=-1$.

\subsection{Power-law expansion}  
The reality condition (\ref{cond4}) becomes:
\begin{equation}
n\left[\,\left(\frac{A}{S}\right)^{\frac{1}{n}}
+ \frac{3}{2}\,\kappa_{0}\,\zeta\,\right]\left(\frac{A}{S}\right)
^{\frac{1}{n}}\,S^{2}- \frac{\Sigma^{2}}{S^{4}} \geq - k\;.
\end{equation}
This relation can be satisfied for both $k=0$ and $k=-1$, subject
to the presence of a lower bound on $S$.

\subsection{$\sinh$--expansion}  
The reality condition (\ref{cond4}) becomes:
\begin{equation}
\label{sinhsbv}
\omega^{2}\,A^{2} - \frac{\Sigma^{2}}{S^{4}} + \frac{3}{2}\,
\kappa_{0}\,\zeta\,\omega\,S\,\sqrt{S^{2}+A^{2}} \geq - k\;.
\end{equation}
Now except for the shear term all other terms on the LHS are positive
and are
either constant or growing with $S$. As a result, both values $k=0$
and $k=-1$ are allowed but there exists a lower bound on $S$ in each
of the two cases.

\subsection{Intermediate $\sinh$--expansion}  
The reality condition (\ref{cond4}) becomes:
\begin{eqnarray}
\label{realsinhesbv}
& &-\epsilon\left(\epsilon-1\right)\omega\,\frac{S\,\sqrt{A^2+S^2}}
{\left[\,\frac{1}{\omega}\,
\mbox{arsinh}\frac{S}{A}\,\right]^{\frac{2-\epsilon}{\epsilon}}}
+ \epsilon^2\omega^2A^2
\left[\,\frac{1}{\omega}\,\mbox{arsinh}\frac{S}{A}\,\right]
^{\frac{2\epsilon-2}{\epsilon}}\\
& &+ \frac{3}{2}\,\kappa_{0}\,\zeta\,\epsilon\,\omega\,S\,
\sqrt{A^{2}+S^{2}}\,\left[\,\frac{1}{\omega}\,
\mbox{arsinh}\frac{S}{A}\,\right]^{\frac{\epsilon-1}{\epsilon}}
- \frac{\Sigma^{2}}{S^{4}} \geq -k\nonumber\;.
\end{eqnarray}
Again, for the sake of simplicity and consistency with previous
sections,
we only analyse (\ref{realsinhesbv}) for values of
the parameter $\epsilon$ around $\epsilon=1$.
Naturally, (\ref{realsinhesbv}) combines the results of
both (\ref{realsinhes}) and (\ref{realsinhebv}), which means that
due to the dominance of the shear term for small values of the
scale factor $S$ has to be constrained from below for both values
of the curvature parameter $k$, whereas depending on the magnitude
of the factor $3/2\,\kappa_{0}\,\zeta\,\omega$ the effects
of the bulk
viscosity term will remove the upper bound on $S$ that was
present in the pure FLRW case of condition (\ref{realsinhe}).

\subsection{Intermediate exponential expansion}  
The reality condition (\ref{cond4}) becomes:
\begin{equation}
\epsilon\,\omega\left[\,\frac{1-\epsilon}{\left[\,\frac{1}{\omega}
\,\mbox{ln}\,\frac{S}{A}\,\right]^{\frac{2-\epsilon}{\epsilon}}}
+ \frac{3}{2}\kappa_{0}\,\zeta\,\frac{1}{\left[\,\frac{1}{\omega}\,
\mbox{ln}\,\frac{S}{A}\,\right]^{\frac{1-\epsilon}{\epsilon}}}
\right]\,S^{2} - \frac{\Sigma^{2}}{S^{4}} \geq - k\;.
\end{equation}
As in the case considered in Section \ref{ssc:iee} and the related
following ones,
the overall situation is very complicated and depends on the values
of the free parameters $\epsilon,~\omega,~A,~\Sigma$ and $\zeta$.
However, for the case of $\epsilon \rightarrow 1$ and sufficiently
large $\omega$, the simultaneous non-zero $\Sigma$ and $\zeta$
perturbations
do not produce any qualitative changes. As a result, in this restricted
sense, the results of Section \ref{ssc:iee} are robust with
respect to such perturbations.

\subsection{de Sitter expansion}  
The reality condition (\ref{cond4}) becomes:
\begin{equation}
- \frac{\Sigma^{2}}{S^{4}} + \frac{3}{2}\,\kappa_{0}\,\zeta\,
\omega\,S^{2}\geq - k\;.
\end{equation}
Now as $S\rightarrow A$ the scale factor $S$ remains unconstrained
for $k=0$ and
$k=-1$ given that $\Sigma \leq \sqrt{\frac{3}{2}\kappa_{0}\,
\zeta\,\omega}\,A^{3}$ and $\Sigma \leq \sqrt{\frac{3}{2}
\kappa_{0}\,\zeta\,\omega\,A^{2}-1}\,A^{2}$ hold respectively.
Otherwise $S$
will be restricted by a lower bound with $S>A$.

\subsection{$\cosh$--expansion}  
The reality condition (\ref{cond4}) becomes:
\begin{equation}
\omega^{2}\,A^{2} + \frac{\Sigma^{2}}{S^{4}} - \frac{3}{2}\,
\kappa_{0}\,\zeta\,\omega\,S\,\sqrt{S^{2}-A^{2}} \leq k\;.
\end{equation}
In this case, for both possible values of the curvature parameter
$k$, a lower
bound on $S$ with $S>A$ has to be imposed.

\subsection{Super-exponential expansion}  
The reality condition (\ref{cond4}) becomes:
\begin{equation}
n\,\omega\left[\,(n-1)\,\left[\,\frac{1}{\omega}\,\mbox{ln}\,
\frac{S}{A}\,\right]^{\frac{n-2}{n}} - \frac{3}{2}\,\kappa_{0}\,
\zeta
\,\left[\,\frac{1}{\omega}\,\mbox{ln}\,\frac{S}{A}\,
\right]^{\frac{n-1}{n}}\,\right]\,S^{2}
+ \frac{\Sigma^{2}}{S^{4}} \leq k\;.
\end{equation}
In this example again we divide the range of the parameter $n$ into
(i) $1<n<2$, (ii) $n=2$ and (iii) $n>2$. In all three regions both
$k=0$ and $k=-1$ are excluded in the limit as $S\rightarrow A$.
For large $S$ the shear term becomes negligible and the situation
becomes qualitatively the same as the one in Section
\ref{sec:bulk}.

\subsection{Double exponential expansion}  
The reality condition (\ref{cond4}) becomes:
\begin{equation}
\omega\left(\omega-\frac{3}{2}\,\kappa_{0}\,\zeta\right)\,S^{2}\,
\mbox{ln}\,\frac{S}{A} + \frac{\Sigma^{2}}{S^{4}} \leq k\;.
\end{equation}
Again, we have to take into account the overall sign of the first
term on the LHS. If $\omega>3/2\,\kappa_{0}\,\zeta$, the
reality condition excludes both cases $k=0$ and $k=-1$. However,
if $\omega<3/2\,\kappa_{0}\,\zeta$, $k=0$ and $k=-1$ will satisfy the
reality condition for large values of $S$ while, depending on the
values of the remaining parameters, there might occur a lower
bound for $S$ as $S\rightarrow A\,\exp B$.

\section{Discussion}
Even though equations (\ref{V}) and (\ref{phidot}) amount to a
correspondence between the spaces of scale factors $\cal{S}$ and
scalar field potentials $\cal{V}$, it is clear that this
correspondence is not free. The positive definiteness of the
kinetic energy density of the scalar field in Eq.
(\ref{phidot}) imposes a reality condition on the set of possible
inflationary models. This constrains the range of scale factors
available and is therefore of potential importance in whether or
not inflation is allowed to take place within a particular model.
Now since the solutions
invoked to account for inflation are not unique, it is important
to see how the degree of restrictiveness imposed by the reality
condition varies for different solutions. To check this, we first
of all confined ourselves to the  standard FLRW framework and
considered
the effect of the reality condition on a number of inflationary
solutions commonly employed in the literature.
Our results show that the degree to which this constraint imposes
restrictions depends on the nature of the solutions under
consideration. Tables 1 and 2 summarise the results of Section
\ref{sec:reality} (note that the abbreviation PD does {\em not}
imply the same parameter dependence in all cases).
As can be seen the effect of the reality
condition is different for both
different solutions and for the same solution with different
parameters.

Now the FLRW framework usually adopted in studies of inflation is
idealised and it would therefore be of value to see what happens to
the results of Section \ref{sec:reality} if small physically
motivated perturbations are considered. In order to achieve this
we

\begin{itemize}
\item firstly considered the effects of including anisotropic
perturbations to the FLRW setting by discussing the
Einstein--Scalar--Field equations on the basis of cosmological
models of Bianchi Type--I and Type--V
(which can be looked upon as anisotropic perturbations to the
geometry of the $k=0$ and $k=-1$ FLRW models respectively).
Tables 3 and 4 summarise the results of Section \ref{sec:shear}
and it is clear that small shear perturbations can have a
significant effect on
some of the cases considered in Section \ref{sec:reality}.
The important point, however, is that the overall effect of the
anisotropy perturbations
is to enhance the prohibitive effects of the reality condition,
as for example can be seen by comparing the $k=0$ rows in the
tables 1 and 3.
\end{itemize}

\begin{itemize}
\item secondly we considered the effects of including viscous
perturbations by including a bulk viscosity term. The results
are summarised
in tables 5 and 6. Again these results indicate that
small viscous perturbations can have important consequences
for the effectiveness of the reality condition. However, as
opposed to the effects of anisotropic perturbations, the overall
effect of the
viscous perturbations
is to decrease the degree of restrictiveness of the
reality condition, as can be seen from the comparison of
$k=-1$ and $k=0$
rows in tables 2 and 6.
\end{itemize}

Now in reality one would expect such perturbations to be present
simultaneously. This is particularly important because
dynamical properties robust with
respect to one particular perturbation may not remain robust if
additional perturbations are switched on (see for example
\cite{acrt92} for a discussion). To study the effects of
simultaneous switching on of perturbations

\begin{itemize}
\item we considered, in Section \ref{sec:all},
the simultaneous effects of the anisotropic
and viscous perturbations.
The results are summarised in tables 7 and 8. What is
observed is that the result of switching on of viscous
perturbations in presence of anisotropic perturbations
is mixed: in some cases the viscous perturbations
decrease the prohibitive effects of the anisotropic perturbations
(as can be seen by comparing the $k=0$ row of the tables
4 and 8), whereas in other cases it does not (as can be seen by
comparing the $k=0$ row of the tables
3 and 7).

\end{itemize}

To summarise, the comparison of tables 1\,-\,8 shows that
above perturbations can produce qualitatively important,
but non-uniform, changes in the restrictive nature
of the reality condition. As a result, solutions excluded
on the basis of the reality condition (for example in the FLRW
setting)
may become permissible when certain perturbations are allowed
(and vice versa). Importantly, the exponential, the
$\sinh$ and the power-law solutions, commonly employed as
inflationary Ans\"{a}tze for the cosmological scale factor of an
FLRW spacetime metric, are seriously affected
in this way.

Another important fact revealed by our results is that the effect
of the reality condition on inflationary solutions with very
similar functional forms can produce qualitatively different
restrictive effects. This can be seen by comparing the
results relating to the exponential
and the intermediate exponential solutions summarised in tables
3 and 4. As can be seen small anisotropic perturbations result
in the exclusion of the exponential solution, whereas the
intermediate exponential solution remains conditionally
allowed. As another example we may consider the cases
(considered in table 1) of
$A\,\sinh \omega t$ and $A\,\sinh(\omega t^{\epsilon})$, where
small deviations of $\epsilon$ from 1 can have important consequences.
This type of behaviour could be of importance in the interpretation of
observations
in order to arrive at a unique inflationary scenario.

\section*{Acknowledgements}
HvE is supported by a Grant from Drapers' Society at QMW;
PKSD thanks the School of Mathematical Sciences, QMW for
hospitality while some of this work was carried out and the FRD
(South Africa) for financial support and RT is supported by
SERC UK Grant number H09454.



\newpage
\begin{table}
\def\tablerule{\noalign{\hrule}}
$$\vbox{\tabskip=0pt \offinterlineskip
\halign to 313pt{
\strut#&\vrule#\tabskip=1em plus 2em&
&#\hfil&\vrule#&#\hfil&\vrule#&#\hfil&\vrule#&#\hfil&\vrule#&#
\hfil&\vrule#&#\hfil&\vrule#
\tabskip=0pt\cr\tablerule
\omit&height2pt&&&&&&&&&&\cr
&&FLRW&&$A\,t^n$&&$A\sinh\omega t$&&$A\sinh\omega t^{\epsilon}$
&&$A\,e^{\omega t^{\epsilon}}$,
$0<\epsilon<1$&\cr
\omit&height2pt&&&&&&&&&&\cr\tablerule
&&$k=-1$&&LB&&$\omega \ge\frac{1}{A}$&& PD&&
PD&\cr\tablerule
&&$k=0$&&Y&&Y&&PD&&Y&\cr\tablerule
&&$k=+1$&&Y&&Y&& PD &&Y&\cr\tablerule}}$$
\caption{Results for the FLRW case (N implies the solution is
excluded by the
reality condition; Y implies the reality condition imposes no
restrictions on the solution; PD implies that the condition can
be satisfied for certain ranges
of parameters;
and LB (or UB) indicate there is a lower (or a upper)
bound on $S$)}
\end{table}
\begin{table}
\def\tablerule{\noalign{\hrule}}
$$\vbox{\tabskip=0pt \offinterlineskip
\halign to 290pt{
\strut#&\vrule#\tabskip=1em plus 2em&
&#\hfil&\vrule#&#\hfil&\vrule#&#\hfil&\vrule#&#\hfil&\vrule#&#
\hfil&\vrule#&#\hfil&\vrule#
\tabskip=0pt\cr\tablerule
\omit&height2pt&&&&&&&&&&\cr
&&FLRW&&$A\,e^{\omega t}$&&$A\cosh\omega t$&&$A\,e^{\omega t^{n}}$,
$n>1$&&$A\,e^{Be^{\omega t}}$&\cr
\omit&height2pt&&&&&&&&&&\cr\tablerule
&&$k=-1$&&N&&N&&N&&N&\cr\tablerule
&&$k=0$&&Y&&N&&N&&N&\cr\tablerule
&&$k=+1$&&Y&&$\omega \le\frac{1}{A}$&& PD&&PD&\cr\tablerule}}$$
\caption{Results for the FLRW case}
\end{table}
\begin{table}
\def\tablerule{\noalign{\hrule}}
$$\vbox{\tabskip=0pt \offinterlineskip
\halign to 341pt{
\strut#&\vrule#\tabskip=1em plus 2em&
&#\hfil&\vrule#&#\hfil&\vrule#&#\hfil&\vrule#&#\hfil&\vrule#&#
\hfil&\vrule#&#\hfil&\vrule#
\tabskip=0pt\cr\tablerule
\omit&height2pt&&&&&&&&&&\cr
&&$\sigma\neq 0$, $\zeta=0$&&$A\,t^n$&&$A\sinh\omega t$&&
$A\sinh\omega t^{\epsilon}$&&$A\,e^{\omega t^{\epsilon}}$,
$0<\epsilon<1$&\cr
\omit&height2pt&&&&&&&&&&\cr\tablerule
&&$k=-1$&&LB&&LB, $\omega >\frac{1}{A}$&& PD&& PD&
\cr\tablerule
&&$k=0$&&LB&&LB&& PD&&LB&\cr\tablerule}}$$
\caption{Results for the $\sigma\neq 0$, $\zeta=0$ case}
\end{table}
\begin{table}
\def\tablerule{\noalign{\hrule}}
$$\vbox{\tabskip=0pt \offinterlineskip
\halign to 312pt{
\strut#&\vrule#\tabskip=1em plus 2em&
&#\hfil&\vrule#&#\hfil&\vrule#&#\hfil&\vrule#&#\hfil&\vrule#&#
\hfil&\vrule#&#\hfil&\vrule#
\tabskip=0pt\cr\tablerule
\omit&height2pt&&&&&&&&&&\cr
&&$\sigma\neq 0$, $\zeta=0$&&
$A\,e^{\omega t}$&&$A\cosh\omega t$&&$A\,e^{\omega t^{n}}$,
$n>1$&&$A\,e^{Be^{\omega t}}$&\cr
\omit&height2pt&&&&&&&&&&\cr\tablerule
&&$k=-1$&&N&&N&&N&&N&\cr\tablerule
&&$k=0$&&N&&N&&N&&N&\cr\tablerule}}$$
\caption{Results for the $\sigma\neq 0$, $\zeta=0$ case}
\end{table}
\begin{table}
\def\tablerule{\noalign{\hrule}}
$$\vbox{\tabskip=0pt \offinterlineskip
\halign to 332pt{
\strut#&\vrule#\tabskip=1em plus 2em&
&#\hfil&\vrule#&#\hfil&\vrule#&#\hfil&\vrule#&#\hfil&\vrule#&#
\hfil&\vrule#&#\hfil&\vrule#
\tabskip=0pt\cr\tablerule
\omit&height2pt&&&&&&&&&&\cr
&&$\zeta\neq 0$, $\sigma=0$&&$A\,t^n$&&$A\sinh\omega t$&&
$A\sinh\omega t^{\epsilon}$&&$A\,e^{\omega t^{\epsilon}}$,
$0<\epsilon<1$&\cr
\omit&height2pt&&&&&&&&&&\cr\tablerule
&&$k=-1$&&LB&&$\omega\ge\frac{1}{A}$&&PD&& PD&\cr\tablerule
&&$k=0$&&Y&&Y&&PD&&Y&\cr\tablerule
&&$k=+1$&&Y&&Y&&PD&&Y&\cr\tablerule}}$$
\caption{Results for the $\zeta\neq 0$, $\sigma=0$ case}
\end{table}
\begin{table}
\def\tablerule{\noalign{\hrule}}
$$\vbox{\tabskip=0pt \offinterlineskip
\halign to 316pt{
\strut#&\vrule#\tabskip=1em plus 2em&
&#\hfil&\vrule#&#\hfil&\vrule#&#\hfil&\vrule#&#\hfil&\vrule#&#
\hfil&\vrule#&#\hfil&\vrule#
\tabskip=0pt\cr\tablerule
\omit&height2pt&&&&&&&&&&\cr
&&$\zeta\neq 0$, $\sigma=0$&&$A\,e^{\omega t}$&&$A\cosh\omega t$
&&$A\,e^{\omega t^{n}}$, $n>1$&&$A\,e^{Be^{\omega t}}$&\cr
\omit&height2pt&&&&&&&&&&\cr\tablerule
&&$k=-1$&&PD&&LB&& PD&&PD &\cr\tablerule
&&$k=0$&&Y&&LB&& PD&& PD&\cr\tablerule
&&$k=+1$&&Y&&$\omega\le\frac{1}{A}$&& PD&& PD&
\cr\tablerule}}$$
\caption{Results for the $\zeta\neq 0$, $\sigma=0$ case}
\end{table}
\begin{table}
\def\tablerule{\noalign{\hrule}}
$$\vbox{\tabskip=0pt \offinterlineskip
\halign to 332pt{
\strut#&\vrule#\tabskip=1em plus 2em&
&#\hfil&\vrule#&#\hfil&\vrule#&#\hfil&\vrule#&#\hfil&\vrule#&#
\hfil&\vrule#&#\hfil&\vrule#
\tabskip=0pt\cr\tablerule
\omit&height2pt&&&&&&&&&&\cr
&&$\sigma\neq 0$, $\zeta\neq 0$&&
$A\,t^n$&&$A\sinh\omega t$&&
$A\sinh\omega t^{\epsilon}$&&$A\,e^{\omega t^{\epsilon}}$,
$0<\epsilon<1$&\cr
\omit&height2pt&&&&&&&&&&\cr\tablerule
&&$k=-1$&&LB&&LB&&PD&&PD&\cr\tablerule
&&$k=0$&&LB&&LB&&PD&&PD&\cr\tablerule}}$$
\caption{Results for the $\sigma\neq 0$, $\zeta\neq 0$ case}
\end{table}
\begin{table}
\def\tablerule{\noalign{\hrule}}
$$\vbox{\tabskip=0pt \offinterlineskip
\halign to 312pt{
\strut#&\vrule#\tabskip=1em plus 2em&
&#\hfil&\vrule#&#\hfil&\vrule#&#\hfil&\vrule#&#\hfil&\vrule#&#
\hfil&\vrule#&#\hfil&\vrule#
\tabskip=0pt\cr\tablerule
\omit&height2pt&&&&&&&&&&\cr
&&$\sigma\neq 0$, $\zeta\neq 0$&&
$A\,e^{\omega t}$&&$A\cosh\omega t$&&$A\,e^{\omega t^{n}}$,
$n>1$&&$A\,e^{Be^{\omega t}}$&\cr
\omit&height2pt&&&&&&&&&&\cr\tablerule
&&$k=-1$&&PD&&LB&&PD&&PD&\cr\tablerule
&&$k=0$&&PD&&LB&&PD&&PD&
\cr\tablerule}}$$
\caption{Results for the $\sigma\neq 0$, $\zeta\neq 0$ case}
\end{table}

\end{document}